\documentclass[english]{aastex}
\usepackage[T1]{fontenc}
\usepackage{float}
\usepackage{amstext}
\usepackage{amssymb}
\usepackage{graphicx}
\usepackage{epstopdf}
\makeatletter
\slugcomment{}
\shorttitle{}
\shortauthors{}

\usepackage{multirow}
\usepackage{rotating}

\makeatother

\usepackage{babel}
\begin{document}

\title{Reducing Uncertainties in the Production of the Gamma Emitting
  Nuclei $^\mathbf{26}\mathbf{Al}$, $^\mathbf{44}\mathbf{Ti}$, and
  $^\mathbf{60}\mathbf{Fe}$ in Core Collapse Supernovae by Using
  Effective Helium Burning Rates}

\author{Sam M. Austin}
\email{austin@nscl.msu.edu}
\affil{National
	Superconducting Cyclotron Laboratory,\\ Michigan State University,
	640 South Shaw Lane, East Lansing, MI 48824-1321, U.S.A.}
\affil{Joint Institute for Nuclear Astrophysics - Center for the Evolution of the Elements}

\author{Christopher~West}
\email{christopher.west@metrostate.edu}
\affil{Minnesota Institute for Astronomy,
	School of Physics and Astronomy,
	The University of Minnesota, Twin Cities,
	Minneapolis, MN 55455-0149, U.S.A.}
\affil{Center for Academic Excellence,
	Metropolitan State University,
	St Paul, MN, 55106, USA}
\affil{Joint Institute for Nuclear Astrophysics - Center for the Evolution of the Elements}

\author{Alexander~Heger}
\email{Alexander.Heger@Monash.edu}
\affil{Monash Centre for Astrophysics,
	School of Mathematical Sciences,
	Monash University, VIC 3800, Australia}
\affil{Minnesota Institute for Astronomy,
	School of Physics and Astronomy,
	The University of Minnesota, Twin Cities,
	Minneapolis, MN 55455-0149, U.S.A.}
\affil{Center for Nuclear Astrophysics, Department of
	Physics and Astronomy, Shanghai Jiao-Tong University, Shanghai
	200240, P. R. China.}
\affil{Joint Institute for Nuclear Astrophysics - Center for the Evolution of the Elements}

\date{\today}

\begin{abstract}
	We have used effective reaction rates (ERR) for the helium burning
	reactions to predict the yield of the gamma-emitting nuclei
	$^{26}\mathrm{Al}$, $^{44}\mathrm{Ti}$, and $^{60}\mathrm{Fe}$ in core
	collapse supernovae.  The variations in the predicted yields for
	values of the reaction rates allowed by the ERR are much smaller than
	obtained previously, and smaller than other uncertainties.  A
	``filter'' for supernova nucleosynthesis yields based on pre-supernova
	structure was used to estimate the effect of failed supernovae on the
	initial mass function-averaged yields; this substantially reduced the
	yields of all these isotopes, but the predicted yield ratio
	$^{60}\mathrm{Fe}/^{26}\mathrm{Al}$ was little affected.  The
	robustness of this ratio is promising for comparison with data, but it
	is larger than observed in nature; possible causes for this
	discrepancy are discussed.
\end{abstract}


\section{Introduction}
Astronomical observations of gamma rays from long-lived radioactive
nuclei provide unique opportunities for nuclear astrophysics.  The
flux of gamma rays from the decay of $^{26}\mathrm{Al}$ can be used to
infer the rate of supernovae (SNe) in the galaxy \citep{Diehl2013}.
And since $^{26}\mathrm{Al}$ and $^{60}\mathrm{Fe}$ are made in
different radial shells of massive stars (e.g., \citealt{Timmes95}),
the ratio of their fluxes can provide a stringent test of massive star
and SN models.  Convincing conclusions, however, require reliable
predictions of the production rate of these gamma emitters in SNe and
the current status is far from satisfactory.

An important problem is the large impact of uncertainties
in the reaction rates $r_{3\alpha}$ and $r_{\alpha,\gamma}$ of the
helium burning reactions: $3\alpha$ and
$^{12}\mathrm{C}(\alpha,\gamma$)$^{16}\mathrm{O}$.  It was found
\citep{Tur2010} that over a range of $\pm 2\sigma$, where $\sigma$ is the
experimental uncertainty in the helium burning rates, the production
of $^{26}\mathrm{Al}$ varies by about a factor of three.  The production
of $^{60}\mathrm{Fe}$ and the ratio of $^{26}\mathrm{Al}$ to
$^{60}\mathrm{Fe}$ vary by much larger factors.  As a result,
predictions of the yields of the gamma nuclei are not robust, and
depend on the particular values of the helium burning rates chosen
from within the allowed experimental ranges.

In this paper, we attempt to address this issue by using an effective
reaction rate (ERR) for the helium burning reactions
\citep{Austin2014,West2013a} to predict the yields of the gamma nuclei.  Compared
to the earlier calculations, this greatly reduces the predicted
variation of their yields.  With the helium burning rate problem then
mainly under control, we examine some of the issues
that remain. In particular, we examine the nature of the effects of
failed supernovae, by considering the model of \citet{OConnor2011}. In this context, we conclude, tentatively, that the ratio of
$^{60}$Fe and $^{26}$Al abundances is a robust observable. Whether this
remains the case when more sophisticated models of these and
related effects is considered appears to remain an unresolved
question.

\section{Method}
This ERR had been determined by parameterizing the two helium burning
rates and fixing the parameters by fitting the results of SN
nucleosynthesis to the abundance pattern \citep{Lodders2010} of isotopes
produced mainly in core-collapse supernovae: the intermediate mass and
\emph{s}-only nuclei.  This procedure simultaneously treats the uncertainties of the two reaction rates in the context of the KEPLER code as described in \citet{Rauscher2002}.  After scaling the rates relative to standard values, as done in \citet{Tur2007}, we found that equivalently good
matches occur along a line correlating the two rates:
$r_{\alpha,\gamma} = r_{3\alpha} + 0.35$ as shown in Fig.~1 of
\citet{Austin2014}.  The line samples the full $\pm 2\sigma$ range of
$r_{3\alpha}$ but $r_{\alpha, \gamma}$ is more constrained; we
therefore plot the results below as a function of $r_{3\alpha}$. We had anticipated that the rates would be constrained in both $r_{\alpha,\gamma}$ and $r_{3\alpha}$, but the fitted production rates did not lead to that constraint.

The yields of the gamma nuclei were obtained by \citet{West2013a} using the KEPLER code \citep{Rauscher2002,Weaver1978,Woosley1995,Woosley2002,Heger2005} to model the
evolution of sets of $12$ initial stellar masses ($12$, $13$, $14$,
$15$, $16$, $17$, $18$, $20$, $22$, $25$, $27$, and $30$
$\mathrm{M}_{\odot}$) from central hydrogen burning to core collapse.
A $1.2 \times{10^{51}}\,$erg explosion was then simulated using a
piston placed at the base of the oxygen shell \citep{Woosley2007}.  For each mass, calculations were made for a rate matrix covering approximately $\pm 2\sigma$
for $r_{\alpha,\gamma}$ and $r_{3 \alpha}$, a total of $176$ rate
pairs.  It is now known that not all massive stars explode (e.g.,
\citealp{Smartt2009}).  To get a rough idea of this effect on yields of the gamma nuclei, we applied a compactness parameter filter \citep{OConnor2011,West2013a,Sukhbold2014},
namely $\xi_{2.5} < 0.25$, to account for these failed SNe.  Stars
with masses $22$, $27$, and $30$ $\mathrm{M}_{\odot}$ as well as a few
$r_{\alpha, \gamma}$, $r_{3\alpha}$ pairs at other masses  did not satisfy this criterion, and were assumed not to
explode.

We then calculated the average yield $Y$ for the Initial Mass Function
(IMF) using the usual formulae:

$$Y_i(m)=\frac{m_{i+1}-m}{m_{i+1}-m_i}Y(m_i)+\frac{m-m_i}{m_{i+1}-m_i}Y(m_{i+1})$$

$$Y=\left[\sum_{i=1}^{N-1}\int_{m_i}^{m_{i+1}}Y(m)\,m^{-2.35}\,\mathrm{d}m\right]/\int_{m_1}^{m_N}m^{-1.35}\,\mathrm{d}m$$

\noindent where $m_i$ and $Y(m_i)$ are taken from the $r_{\alpha, \gamma}$ versus  $r_{3\alpha}$ grid \citep{West2013a,Austin2014}.

\section{Results and Discussion}
\begin{figure}
	\centering \includegraphics[width=\columnwidth]{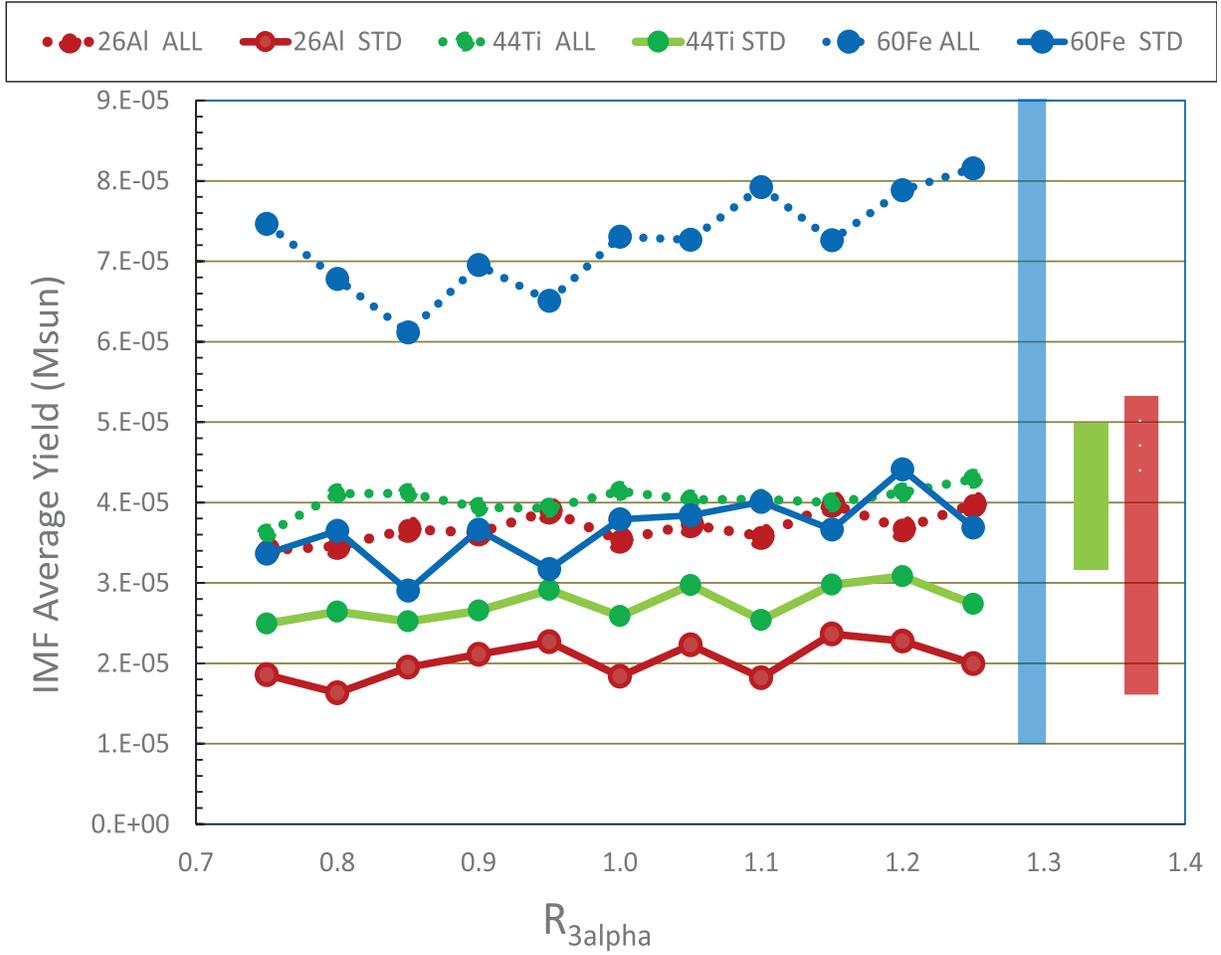}
	\caption{(color online) IMF averaged yields of $^{26}\mathrm{Al}$,
		$^{44}\mathrm{Ti}$, and $^{60}\mathrm{Fe}$ along the ERR line for
		the STD and ALL cases.  The vertical bars (to be compared to the dotted curves) are variations  
		found in previous calculations \citep{Tur2010} for the
		\citet{Lodders2003} abundances; for $^{60}\mathrm{Fe}$ the bar extends to
		$17\times{10^{-5}}$.
		\label{fig:yield}}
\end{figure}

In Fig.~\ref{fig:yield}, we show the results of these calculations,
expressed as an average over a Salpeter IMF with an exponent of
$-2.35$.  The results are given for equally spaced (in $r_{3\alpha}$)
points along the ERR line. For our standard case, labeled STD we omit
the explosive yields of the failed supernovae, but include wind
contributions since winds are mainly emitted before the onset of core
collapse.  To show the effects of the compactness parameter filter, we also give
the results of the unfiltered calculations, including the yields for
all calculated stars (labeled ALL).  As expected the IMF averages for
the $M_{sun}\leq 20$ subset of our grid (not shown) differ little from
the STD case.

In this figure, the mass-to-mass variations arise mainly from binning
effects; not all simulations were performed at points that lay
precisely on the ERR line and some interpolation was required.

There are two immediate conclusions from this figure.  First, the
yield variations are rather small for allowed helium burning rates,
those on the ERR line.  This is true for both the STD and ALL
results.  In contrast, the variations corresponding to independent
uncertainties in $r_{3\alpha}$ and $r_{\alpha,\gamma}$, as determined
in \citet{Tur2010} are much larger as shown by the
colored bars.  And second, the effects of the simple compactness parameter filter are
rather large, as shown by the differences of the STD and ALL results.

\begin{figure}[h]
	\centering
	\includegraphics[width=\columnwidth]{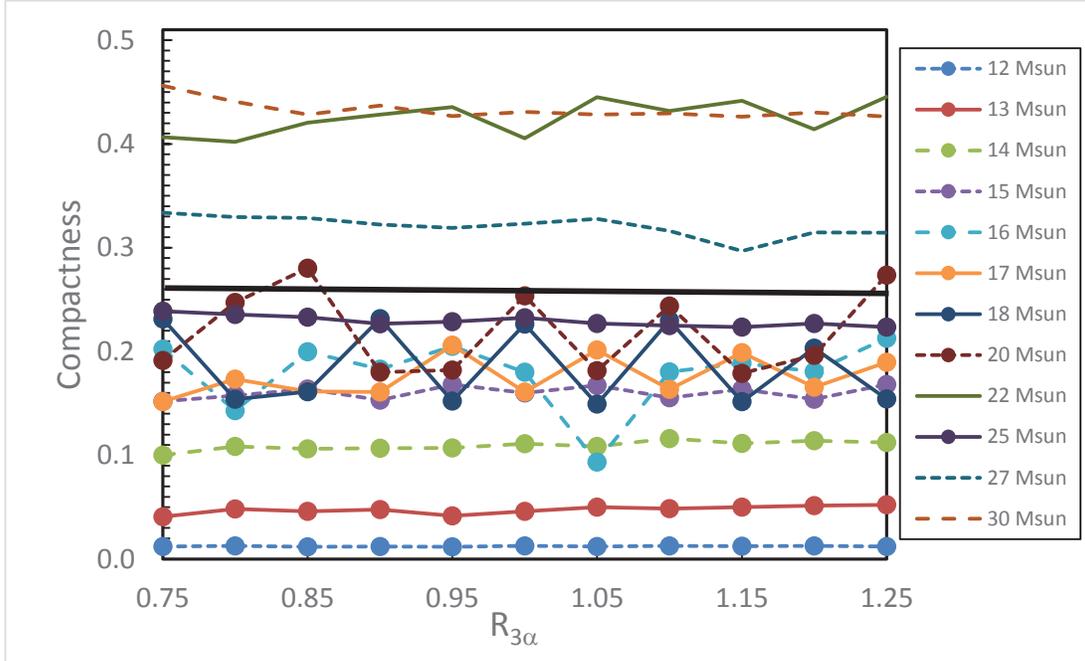}
	\caption{(color online) Compactness parameters. The heavy dark line at
		$\xi_{2.5}=0.25$ is the value assumed to divide stars that are
		likely to explode in the model of \citet*{OConnor2011}, from those that are not.
		\label{fig:compactness}}
\end{figure}

Given the importance of the compactness parameter filter, we show in
Fig.~\ref{fig:compactness} how $\xi_{2.5}$ depends on mass and
reaction rates along the ERR line.  Previously, there have only been
estimates, see for example, \citep{Sukhbold2014} for variations in $r_{\alpha\gamma}$.  The
results for our grid indicate that if the rates are varied along the
ERR line, variations of $\xi_{2.5}$ are relatively small for most
stars.  The $20\,\mathrm{M}_{\odot}$ star shows larger
fluctuations, perhaps related to its complex convection structure
\citep{Rauscher2002,Limongi2006,Tur2010}. See \citet{Sukhbold2014} for
a detailed discussion.

\begin{figure}
	\centering
	\includegraphics[width=\columnwidth]{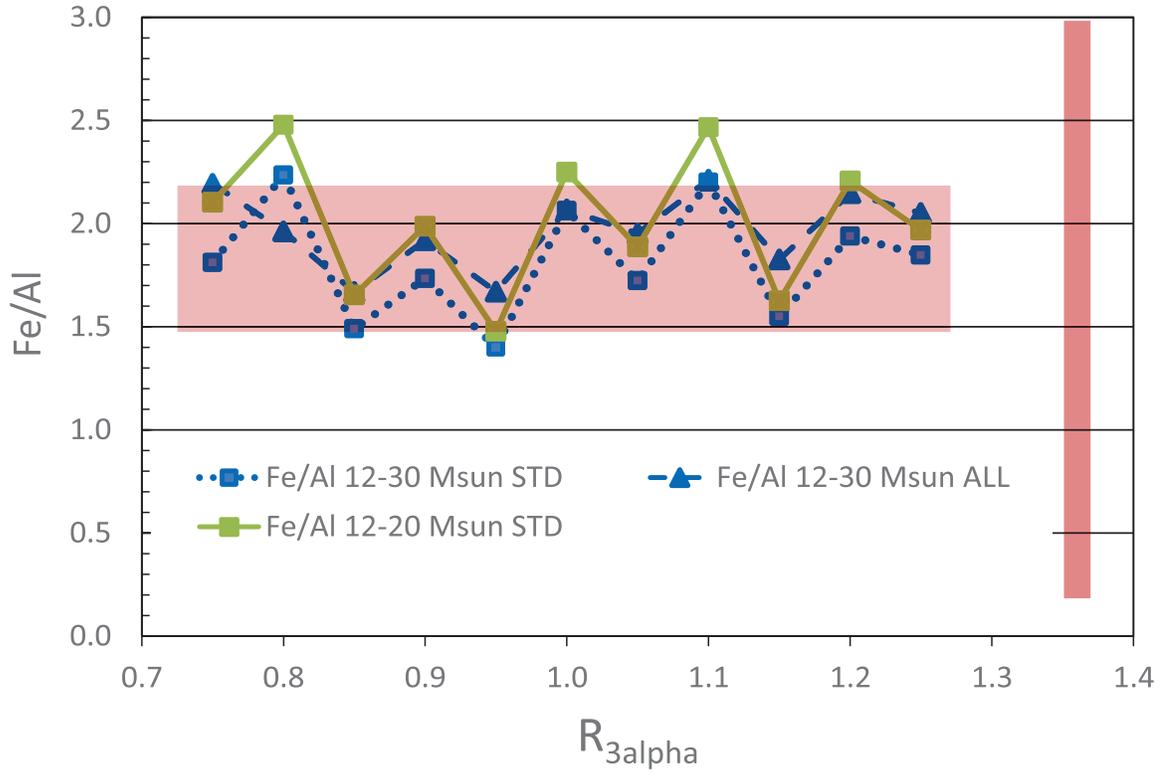}
	\caption{(color online) Ratio $\mathrm{Fe}/\mathrm{Al}$ of the IMF averaged yields
		for the STD and ALL results of Fig.~\ref{fig:yield}, and for the
		range $12-20\,\mathrm{M}_{\odot}$. The horizontal band covering
		$\pm 18\,\%$ gives an indication of the precision of the present
		results.  The narrow vertical bar shows the variations found in
		previous calculations \citep{Tur2010} for the \citet{Lodders2003} abundances; the bar extends to $10$.
		\label{fig:ratio}}
\end{figure}

Results for the $\mathrm{Fe}/\mathrm{Al}$ production ratios are shown
in Fig.~\ref{fig:ratio}.   Again the use of the ERR significantly reduces the
variations compared to those obtained earlier \citep{Tur2010}.  The
interpolation effects are relatively small, and the ratios are
rather similar for the STD, ALL and $12$-$20\,\mathrm{M}_\odot$ cases.

All these values, however, are significantly larger than the values
($\pm1\sigma$ range) of $0.20-0.46$ inferred from the flux
observations \citep{Wang2007,Diehl2013} by multiplying them by $60/26$. A recent paper describing flux observations in detail \citep{Bouchet2015} yields entirely consistent results
with similar uncertainties.

The present predictions for this
ratio and for the individual yields for given stellar masses are also
larger than those of \citet{Limongi2006} (LC); see \citet*{Tur2010}.  One
might speculate that this is due to different treatments of convection
and to other stellar model choices that affect the details of the
convection structures of a star.  Convective processes can, for
example, carry nuclei to hotter regions of the star where their
effective life time and survival probability are significantly
reduced.  These effects can be large for the gamma nuclei or other
nuclei involved in their production ($^{59}\mathrm{Ni}$, for example)
as discussed in \citet{Tur2007,Tur2010}.  Alternatively, these
differences may arise from different choices for the helium burning
rates which can also affect the convection structure of the star.  As
noted above these effects can be large.

It is encouraging that the more
detailed approach of \citep{Sukhbold2016}, apparently using the
same rates for non-helium burning reactions as in this letter,
leads to a decrease in the Fe/Al ratio.  However, Sukhbold did not
consider the effects of uncertainties in the two helium burning
reactions discussed here, and in \citep{Tur2010}, so this
conclusion is tentative.

Differences in other reaction rates also contribute. The
present simulations are part of an extended series of simulations
\citep{Tur2007,Tur2009,Tur2010,West2013a} aimed at understanding the effects of
uncertainties in the helium burning rates on various observables.  For
this purpose we chose to use the default KEPLER rates for other
reactions, even though some had been superseded.
\citet{Woosley2007} (WH) and \citet{Brown2013} (BW) discussed the
effects of updating these rates and found that the ratio was reduced
to about $1.0$.  These authors also discuss other changes in reaction
rates, in explosion energies, and in stellar models that would produce
further effects.  The most important changes were to update the rates
for the $^{26}\mathrm{Al}(\mathrm{p}, \mathrm{n})^{26}\mathrm{Mg}$ and
$^{26}\mathrm{Al}(\mathrm{n},\alpha)^{23}\mathrm{Al}$ reactions, but a
final resolution of these issues will probably require additional
measurements \citep{Iliadis2011}.  Changes in the opacities used in certain
regions of the star were also important.

It is also possible that there are other sources of $\mathrm{Fe}$ or
$\mathrm{Al}$.  The galactic mass of $^{26}\mathrm{Al}$ is
$1.5-3.6\,\mathrm{M}_{\odot}$ \cite{Diehl2013}.
\citet{Bennett2013} and \citet{Wrede2014} note that up to $0.6\,\mathrm{M}_{\odot}$ of
galactic $^{26}\mathrm{Al}$ could be produced by classical novae.
This would increase the ratio in the contributions of massive stars,
but not by enough to remove the discrepancy.

The LC, WH, and BW calculations include contributions from stellar
masses above $30\mathrm{M}_{\odot}$.  It is not clear, however, to
what extent these masses are relevant.  The estimates of \citep{Sukhbold2016} indicate that most stars with $M>30\,\mathrm{M}_{\odot}$ do not explode, although they may expel most or all of their envelope. Other newer simulations \citep{Pejcha2015,Ertl2016,Sukhbold2016,Muller2016,Cote2016}, also allow explosions for larger masses in some cases.  Characterization of a
complex phenomenon in terms of a single compactness parameter is a
substantial approximation, and the newer simulations indicate that
more complex criteria yield a sharper distinction between explosive
and non-explosive scenarios.  It seems a safe conclusion, however, that
much remains to be done before this issue is settled.

There remains the issue of the ERR itself.  Once one has determined
such an effective rate the principal test is that it reproduces a
variety of observables not involved in its determination.  So far we
have shown that using the ERR, rather than the central values of the
rates with errors treated as independent, greatly reduces
variations owing to uncertainties in the helium burning reaction rates
for: the values of the central carbon fraction at the end of helium
burning and of the remnant mass \citep{West2013a}; the yields of the
neutrino nuclei \citep{Austin2014}; and in this paper, the yields of the gamma nuclei.  This satisfies an important necessary condition, but
there remains the question of whether the absolute values of the
observables are reproduced. Since one does not know any of the observed or predicted values with the necessary accuracy it is perhaps useful in
this circumstance to estimate the yield changes owing to the uncertainty  in our determination of the ERR.

One can obtain an estimate of the uncertainty in the location of the
ERR line from the detailed discussion in
\citet{West2013a} where the location is specified by $r_{\alpha,\gamma} =
r_{3\alpha} + b$ and $b=0.35 \pm 0.2$; our calculations used $b=0.35$.
We have repeated the calculations for $b=0.2$ and $b=0.5$.  We find
that the average differences in yields, compared to those for the
central value, are $18\,\%$, $7\,\%$, and $22\,\%$ for
$^{26}\mathrm{Al}$, $^{44}\mathrm{Ti}$, and $^{60}\mathrm{Fe}$,
respectively.  For $^{26}\mathrm{Al}$ and $^{44}\mathrm{Ti}$ the
deviations are largest toward $b=0.5$ and for $^{60}\mathrm{Fe}$
toward $b=0.2$.

It appears that uncertainties in the ERR for helium burning
reactions introduce yield uncertainties that are smaller than those
resulting from other uncertainties.  The uncertainties arising from
the determination of which stars explode are perhaps the largest.

\section{Conclusions}
We find that:

(1) Using the ERR for the helium burning reactions, rather than
treating the rates and their uncertainties as independent, results in
much smaller variations in predicted $^{26}\mathrm{Al}$ and
$^{60}\mathrm{Fe}$ yields and their ratio in supernovae, as is shown
in Figs.~\ref{fig:yield} and ~\ref{fig:ratio}.  The variations are
smaller than other uncertainties.

(2) The $^{60}\mathrm{Fe}/^{26}\mathrm{Al}$ yield ratio may be the
most robust observable involving the gamma nuclei.  Systematic
observational errors are smaller for the ratio than for individual
yields.  We have shown that predictions of the ratio do not depend
strongly on the helium burning rates or on the sample of stars
considered, or on which stars undergo successful explosions.  Given
the present uncertainty in this latter determination this is an
important advantage. Other mechanisms may
eject part of the envelop in weak and/or failed supernovae and lead to
additional 26 Al production; see \citet{love2013} for a theoretical description and \citet{adams2016} for observational evidence.

(3) Use of the ERR may provide a superior approach to reducing the
uncertainties in nucleosynthesis yields due to uncertainties in
convective structure and boundary mixing during core helium burning.
The strong yield variations in the earlier results, especially for
$^{60}\mathrm{Fe}$, were ascribed to the sensitivity of the convection
structure of the star to the helium burning rates \citep{Rauscher2002,Tur2010}.

(4) Unfortunately, we cannot at present take advantage of the
transparency of the galaxy to high energy gamma rays and the accurate
high resolution observations from the SPI spectrometer on the INTEGRAL
satellite \citep{Diehl2013}.  Other relevant reaction rates and simulation inputs
need to be improved. In addition to the uncertainties in the fraction of supernovae that
explode, there remain, for example, questions on the effects of
Wolf-Rayet winds on the production of 26Al, the effects of the
explosion energy on explosive burning yields, changes arising when
evolving stars are part of a binary system, and effects and
uncertainties in the convection structure of evolving stars.

Research support from: US NSF; grants PHY08-22648 (JINA), PHY-1430152
(JINA-CEE), PHY11-02511; US DOE: contract DE-AC52-06NA25396, grants
DE-FC02-01ER41176, FC02-09ER41618 (SciDAC), DE-FG02-87ER40328.  AH was
supported by an ARC FutureFellowship (FT120100363) and SMA by Michigan
State University.


\begin{thebibliography}{}

\bibitem[Adams et al.(2016)]{adams2016} Adams, S. M., Kochanek, C. S.,
Gerke, J. R.,\& Stanek, K. Z., 2016,arXiv:1610.02402

\bibitem[Austin et al.(2014)]{Austin2014}Austin, S., West, C., \& Heger, A., 2010, Phys. Rev. Lett., \textbf{112}, 11

\bibitem[Bennett et al.(2013)]{Bennett2013}Bennett, M., Wrede, C., Chipps, A., Jose, J., Liddick, M., et al., 2013, Phys. Rev. Lett., \textbf{111}, 23

\bibitem[Bouchet et al.(2015)]{Bouchet2015}Bouchet, L., Jourdain, E.,
\& Rooques, J.-P., 2015, ApJ, \textbf{801}, 142

\bibitem[Brown and Woosley(2013)]{Brown2013}Brown, J., \& Woosley, S., Nucleosynthetic Constraints on the Mass of the Heaviest Supernovae. ArXiv e-prints, February 2013.

\bibitem[Cote et al.(2016)]{Cote2016}Cote, B., West, C., Heger, A., Ritter, C., O'Shea, B., et al., 2016, MNRAS, \textbf{463}, 4

\bibitem[Diehl (2013)]{Diehl2013}Diehl, R.,2013, Reports on Progress in Physics, \textbf{76}, 2

\bibitem[Ertl et al.(2016)]{Ertl2016}Ertl, T., Janka, H., Woolsey, S., Sukhbold, T., \& Ugliano, M., 2016, ApJ, \textbf{818}, 124

\bibitem[Heger et al.(2005)]{Heger2005}Heger, A., Kolbe, E., Haxton, W., Laganke, G., Martinez-Pinedo, G., \& Woosley, S., 2005, Phys. Lett. B, \textbf{606}, 258

\bibitem[Iliadis et al.(2011)]{Iliadis2011}Iliadis, C., Champagne, A., Chieffi, A., \& Limongi, M., 2011, ApJs \textbf{193}, 16

\bibitem[Limongi and Chieffi(2006)]{Limongi2006}Limongi, M., \& Chieffi, A., 2006, ApJ, \textbf{647}, 483

\bibitem[Lodders(2003)]{Lodders2003}Lodders, K., 2003, ApJ, \textbf{591}, 1220

\bibitem[Lodders(2010)]{Lodders2010}Lodders, K., 2010, Principles and Perspectives in Cosmochemistry, page 379

\bibitem[Lovegrove and Woosley (2013)]{love2013}Lovegrove, E., Woosley, S. E., 2013, ApJ, \textbf{769}, 109

\bibitem[Muller et al.(2016)]{Muller2016}Muller, B., Heger, A., Liptai, D., \& Cameron, J., 2016, MNRAS, \textbf{460}, 742

\bibitem[O'Connor and Ott(2011)]{OConnor2011}O'Connor, E., Ott, C.,
2011, ApJ, \textbf{730}, 70

\bibitem[Pejcha and Thompson(2015)]{Pejcha2015}Pejcha, O., \& Thompson, T., 2015, ApJ, \textbf{801}, 90

\bibitem[Rauscher et al.(2002)]{Rauscher2002}Rauscher, T., Heger,
A., Hoffman, R., \& Woosley, S., 2002, ApJ, \textbf{576}, 323348

\bibitem[Smartt(2009)]{Smartt2009}Smartt, S., 2009, Annu. Rev. Astron. Astrophys., \textbf{47}, 106

\bibitem[Sukhbold and Woosley(2014)]{Sukhbold2014}Sukhbold, T., \& Woosley, S., 2014, ApJ, \textbf{783}, 10

\bibitem[Sukhbold et al.(2016)]{Sukhbold2016}Sukhbold, T., Ertl, T., Woosley, S., Brown, J., \& Janka, H., 2016, ApJ, \textbf{821}, 38

\bibitem[Timmes et al.(1995)]{Timmes95} Timmes, F. X., Woosley, S. E., Hartmann, D. H., Hoffman, R. D., Weaver, T. A.; Matteucci, F., 1995, ApJ,  \textbf{449}, 204

\bibitem[Tur et al.(2007)]{Tur2007}Tur, C., Heger, A., \& Austin,
S., 2007, ApJ, \textbf{671}, 821

\bibitem[Tur et al.(2009)]{Tur2009}Tur, C., Heger, A., \& Austin,
S., 2009, ApJ, \textbf{702}, 1068

\bibitem[Tur et al.(2010)]{Tur2010}Tur, C., Heger, A., \& Austin, S., 2010, ApJ, \textbf{718}, 357

\bibitem[Wang et al.(2007)]{Wang2007}Wang, W., Harris, M., Diehl, R., Hallion, H., Cordier, B., et al., 2007, A\&A, \textbf{469}, 1005

\bibitem[Weaver et al.(1978)]{Weaver1978}Weaver, T., Zimmerman, G.,
\& Woosley, S., 1978, ApJ, \textbf{225}, 1021

\bibitem[West et al.(2013)]{West2013a}West, C., Heger, A., \& Austin, S., 2013, ApJ, \textbf{769}, 2

\bibitem[Woosley and Weaver(1995)]{Woosley1995}Woosley, S., Weaver,
T., 1995, ApJs, \textbf{101}, 181

\bibitem[Woosley et al.(2002)]{Woosley2002}Woosley, S., Heger, A.,
\& Weaver, T., 2002, Rev. of Mod. Phys., \textbf{74}, 1015

\bibitem[Woosley and Heger(2007)]{Woosley2007}Woosley, S., \& Heger,
A., 2007, Phys. Rep. \textbf{442}, 269

\bibitem[Wrede(2014)]{Wrede2014}Wrede, C., 2014, Personal communication

\end{thebibliography}
\end{document}